# Coupling and induced depinning of magnetic domain walls in adjacent spin valve nanotracks

## Authors

J. Sampaio, L. O'Brien, D. Petit, D.E. Read, E.R. Lewis, H.T. Zeng, L. Thevenard, S. Cardoso, R.P. Cowburn

**Affiliations**

| | |
|---|---|
| J. Sampaio | UMR 137 CNRS-Thales, Palaiseau, France |
| L. O'Brien | University of Minnesota, MN, USA; Thin Film Magnetism Group, Cavendish Laboratory, University of Cambridge, UK |
| D.E. Read | University of Cardiff, UK |
| E.R. Lewis, H.T. Zeng | Physics Department, Imperial College London, UK |
| D. Petit, R.P. Cowburn | Thin Film Magnetism Group, Cavendish Laboratory, University of Cambridge, UK |
| L. Thevenard | Institut des Nanosciences de Paris, UMR 7588, France |
| S. Cardoso | INESC-MN/Institute for Nanosciences and Nanotechnologies, Lisbon, Portugal; Physics Department, IST, Lisbon, Portugal |

**Corresponding author:** J. Sampaio `joao-miguel.sampaio@thalesgroup.com` .

## Abstract

The magnetostatic interaction between magnetic domain walls (DWs) in adjacent nanotracks has been shown to produce strong inter-DW coupling and mutual pinning. In this paper, we have used electrical measurements of adjacent spin-valve nanotracks to follow the positions of interacting DWs. We show that the magnetostatic interaction between DWs causes not only mutual pinning, as observed till now, but that a travelling DW can also induce the depinning of DWs in near-by tracks. These effects may have great implications for some proposed high density magnetic devices (e.g. racetrack memory, DW logic circuits, or DW-based MRAM).

**Keywords:** magnetic domain walls, magnetization reversal mechanisms, giant magnetoresistance, spin valves

**PACS:** 75.47.De, 75.60.Ch, 75.60.Jk, 75.70.Cn, 75.75.Cd, 75.78.Cd, 85.70.Kh, 85.75.-d

The magnetostatic interaction between domain walls (DWs) in adjacent nanotracks has been shown to cause the DWs to couple, pin strongly, and even to change their structure[1–3,16]. This interaction is of interest to fundamental research of pinned DWs, as the coupled pairs of transverse DWs (TDWs) form a well-defined pinning potential with negligible DW deformation[1,2]. Furthermore, it has wide implications for technological applications of magnetic nanotracks, introducing density limitations in proposed DW-based logic and data storage devices[4,5]. Finally, it can be used in new devices, either as a simple configurable DW pinning mechanism or as the basis of a resonator[6,16]. So far, the inter-DW magnetostatic interaction has been observed in monolayer nanotracks via its pinning effects[1] or via its modulation of the structure of static DWs[3,16]. In this paper, we demonstrate experimentally the magnetostatic interaction between DWs in the free layers of adjacent spin valve (SV) nanotracks. The interest in SV track systems arises from the possibility to measure electrically the DW position using the giant magnetoresistance (GMR) effect[7], and from its application to the electronic integration of DW-based spintronic devices. We observed inter-DW pinning and formation of coupled DW pairs, as well as the depinning of a DW by a near-by travelling DW, a phenomenon unobserved up to now.

The SV stack was deposited on a Si/SiO$_2$ substrate by ion-beam deposition in a Nordiko3000 tool[8], with a Permalloy (Py = Ni$_{81}$Fe$_{19}$) free layer and a synthetic anti-ferromagnet reference layer (Ta 2 nm/ Py 8 nm/ Co$_{80}$Fe$_{20}$ 2 nm/ Cu 2 nm/ Co$_{80}$Fe$_{20}$ 2 nm/ Ru 0.8 nm/ Co$_{80}$Fe$_{20}$ 2 nm/ Mn$_{76}$Ir$_{24}$ 6 nm/ Ta 5 nm). It was patterned by ion etching through a Ti hard mask, which itself was patterned by electron-beam lithography and lift-off. An SEM image of the structure is shown in **Fig. 1**. It consisted of two 260nm wide tracks, one L-shaped (top) and the other U-shaped (bottom), placed in close proximity (gap = 70 nm). The reference layer was pinned in the horizontal direction (-X). Three Ti/Au contacts were patterned by a separate lift-off step, one a common ground contact plus an individual contact to each SV track. The SV horizontal segments were both 4.7 µm long, though the sections between the contacts were 1.7 (L track) and 2.4 µm long (U track). The resistance versus time of the two parallel sections of the SV tracks were measured simultaneously using independent lock-in amplifiers with separate AC current sources (measurement current was 10 µA, corresponding to j ≈ 0.5 MA·cm$^{-2}$, frequency from 4 to 40 kHz). The applied current density is too low to induce significant spin transfer torques on the DW [4]. Due to signal filtering, the time resolution was ~0.5 ms, orders of magnitude above the expected DW transit times (expected to be of the order of few to hundreds of ns). For clarity, the normalized resistance was used, MR(t) = ( R(t) – $R_0$ ) / $R_0$ , where R(t) is the resistance and $R_0$ the lowest resistance value (obtained with the magnetization of the free and reference layers parallel). By applying a magnetic field sequence with an in-plane quadrupole electromagnet, transverse DWs were injected and propagated in the free layer of the SV tracks as follows (field sequence is shown in the inset of **Fig. 2**). First, a reset field of 560 Oe ($H_{RESET}$) in the (-X, -Y) direction was applied, magnetizing the free layer of horizontal track segments leftwards and of vertical segments downwards, and initializing a head-to-head TDW in the corner of the L track and a tail-to-tail TDW in the right corner of the U track. The expected DW structure is of the asymmetric transverse type even though a vortex DW would be a lower energy state for the used track dimensions[17]. This is due to the initialization procedure: a large external field applied to a curved track stabilises TDWs. These are stable even under the application of external fields and in the presence of pinning sites (as frequently observed in similar experiments[18,19].) Additionally, the inter-DW interaction is expected to stabilize the TDW[1]. The DWs were then pushed in opposite directions by a ramping horizontal field (0 to +200 Oe, under constant $H_Y$ = -20 Oe), which forced the DWs to pass in close proximity.

The MR levels versus the sweeping $H_X$ (top axis) and time (bottom axis) are shown in the top two curves of **Fig. 2**. This is a simultaneous non-averaged measurement of both tracks. The field precision, limited by the measurement rate, is 0.4 Oe. It can be seen that the magnetization reversal of both tracks occurs in multiple steps at field values between 10-50 Oe. This is due to natural pinning centers in the tracks. The field required for nucleation of new domains, measured separately, is much higher (200 Oe), ruling out the injection of new DWs.

The MR signal of each track is proportional to the ratio of the leftwards to rightwards domains in the section between the contacts. As the SV tracks are in a single DW state, the MR level can be linearly mapped to the position of the DW between the contacts[7], denoted by the variable *d* measured from the ground contact(see **Fig. 1**). For example, for the L track with a head-to-head DW, the MR level 0% corresponds to *d*=1.7 µm (when the DW is in the corner and the free layer is magnetized leftwards) and the maximum MR level (2%) corresponds to *d*=0 µm (when the DW has gone past the common contact and the free layer is magnetized rightwards).

The calculated position *d* as a function of time (bottom axis) and $H_X$ (top axis) is shown in the bottom panel of **Fig. 2** for both tracks. The noise level was equivalent to an error in *d* of ±60 nm. In this plot, we see that the DW in the U track first depins from its corner at 12 Oe, is pinned at some natural defect at $d$=0.43 µm, until at 19 Oe it travels to $d$=1.58 µm where it meets the DW in the L track, which is still close to its initial corner. The two DWs stay pinned at the same $d$ position ($d$=1.58 µm), until they finally decouple at the same value of field of 49 ±0.4 Oe. While pinned, the positions of the DWs are virtually identical, differing initially only by 0.01±0.02 µm (from $H_X$ = 18 to 42 Oe), and by 0.15 µm just before decoupling. This difference is below the accuracy of the measurement, and smaller than the TDW lateral extension (which, for the TDW'S largest side, is similar to the track width[9], 260 nm). This experiment was repeated on the same structure, yielding similar observations, though with some variation of the field at which the DW pair simultaneously depin (ranging 36–49 Oe in 10 measurements). Such large variation in depinning fields is commonly observed in SV tracks, though not completely understood[10–13]. This coincidence in pinning position and depinning field of the DW pair in multiple observations strongly supports the hypothesis of DW pair formation and DW induced pinning, with a decoupling field of 42±6 Oe. Additionally, similar observations were obtained in nominally identical structures, with the coupled DW pair sitting at various $d$ positions and with different decoupling field values, as would be expected given sample to sample variations in inter-track gap and pinning defects[1,2]. The magnitude of the decoupling field further confirms the previous hypothesis that the coupled DWs are of the transverse type and coupled by their wide side, as any other configuration would yield a much smaller coupling (see discussion in references 1 and 16). This coupled configuration, however, does not rule out the precessional DW reversal during propagation associated with propagation above the Walker field[15]. For structures with larger gaps (>130 nm), simultaneous depinning of DWs from the same position was not observed, due to the fact that DW coupling is much weaker for larger gaps[1], becoming negligible in face of the natural pinning defects.

This system also revealed that DW-DW interaction has effects other than DW induced pinning and coupled pair formation. In some structures, a travelling DW was observed to cause the other DW, pinned at some natural defect, to depin without the formation of a static DW pair. This case can be seen in the measurement of **Fig. 3** (similar structure as previously, track separation 110 nm). There, the two DWs propagated through multiple (uncorrelated) propagation-pinning events between $H_X$ = 0–30 Oe. At $H_X$ = 30 Oe, both DWs depinned within the time resolution of the setup from pinning centers 0.67 µm apart. The large separation between the initial DW positions rules out the possibility that the dws were pinned by each other. As in the previous case, the coincidence in depinning fields strongly suggests that these events were correlated, as does the repeated observation of this occurrence in the same structure (we observed this phenomenon in 11 out of 20 measurements) and in other identical structures (in a group of 6 structures, we observed this in 4). We propose that firstly, at $H_X$ = 30 Oe, one of the DWs depinned and travelled towards the end of the track, passing by the other DW while not being pinned by it. The perturbation caused by the magnetostatic field of the travelling DW induced the depinning of the still-pinned DW, which then travelled to the end of its track. The two DW movements appear simultaneous as the DW travel times are much faster than the time resolution of the measurement (tens of ns versus 5 ms). Which of the two DWs moved first cannot be determined by this measurement. This can be seen as a related case to the previous, with the difference that, before crossing, both DWs were pinned with a pinning field greater than the dynamic coupling field of the DW pair. In other words, this phenomenon is observed in tracks where the depinning field from natural defects is comparable or greater than the DW pair coupling field.

To further understand this depinning mechanism, we conducted micromagnetic simulations[14] to reproduce the perturbation of a travelling DW on a pinned DW. We simulated two parallel Permalloy nanotracks ($M_S$=8·10$^5$ A/m, A=13·10$^{-12}$ J/m, α=0.01, T=0 K) with 260×8 nm$^2$ cross-section, 3 μm length, and a 110 nm inter-track gap (cell size was 5×5×8 nm$^3$). The two tracks were initialized with two asymmetric transverse DWs of opposite polarity, 1.6 μm apart (the initial magnetization is shown in the first inset of **Fig. 4**). The static decoupling field of a DW pair, simulated separately, was 29.0 Oe. The top DW was pinned at a 20 nm deep triangular notch ($H_{DEPIN}$ = 46 Oe, simulated separately), while the bottom DW was free. **Fig 4** shows the positions of the DWs during the application of an external field $H_X$=29 Oe. Initially, the top DW was static while the bottom DW propagated forward. As the travelling DW approached the pinned DW, the mutual attraction caused the latter to depin towards the incoming DW (second inset of **Fig. 4**), even though the applied field was only ~60% of the unaided depinning field of the notch (46 Oe). This occurred when the two DWs were ~0.50 μm apart (about 2 DW widths). With the top DW free from the notch, the DWs then decoupled and propagated separately towards opposite ends of the tracks (third inset of **Fig. 4**). During propagation, and also while the DWs were close together, the DWs showed oscillating velocities and cyclical structural transformations (TDW to anti-vortex to reversed TDW), as is typical of the Walker breakdown process[15]. These DW structure transformations strongly modulate the DW coupling strength[1], and may potentially prevent the induced depinning process. However, various simulations with larger values $H_0$, and thus different Walker reversal periods, showed similar induced depinning. This can be understood by the fact that, in all performed simulations, the traveling DW presented the strongest coupling structure (TDW with wide side up) at some point within the induced depinning range (which is, as seen before, at least ~0.50 µm). Simulations at lower $H_X$ showed either DW pair formation ($H_0$< 10 Oe) or propagation of the bottom DW without induced depinning of the top DW (10 < $H_0$ < 27 Oe) [20].

These simulations shows that a travelling DW can induce the depinning of a DW in a separate track, at an external field significantly lower than the unaided depinning field. It also shows that this effect is not suppressed by the DW structure reversals of Walker Breakdown regime. Finally, it supports the hypothesis proposed for the observations shown in **Fig. 3**.

This work shows that a travelling near-by dw can be used to generate fast-varying local magnetic fields that affect pinned DWs, greatly lowering their depinning fields. Furthermore, as it is mediated by the magnetostatic field of the DW, this effect should be equally active in structures with other DW driving forces (such as current induced DW propagation), as well as in structures with different geometries (such as vertically piled tracks). The local nature of these effects could render them of interest to device applications where selective or controlled DW depinning is needed, such as proposed DW-based logic and data storage devices[4,5], and possibly at the same time lowering the current or field needed to depin these DWs.

Concluding, we have simultaneously measured the position of DWS in the free layer of adjacent SV nanotracks, using quasi-static GMR measurements. This allowed us to observe inter-DW pinning, the creation of coupled DW pairs, and the induced depinning effect of a near-by travelling DW, an effect so far unobserved. In order to better understand this last effect, we also carried out micromagnetic simulations that were consistent with the induced depinning hypothesis. This work shows that SV tracks are a useful system to study inter-DW interactions, and suggests that these interactions may have further effects so far unknown. It also shows that a travelling near-by DW can be used to generate fast-varying local magnetic fields that affect pinned DWs, which could be of interest to device applications where selective DW depinning is needed. All these observations in SV tracks should also be valid in single layer tracks. We believe that the measurement of DW positions at faster time-scales (demonstrated in SV tracks in reference [12], among others) might bring a clearer light to these interactions.

# Acknowledgements

This work was supported by the European Community under the Sixth Framework Programme SPINSWITCH MRTN-CT-2006-035327.

# Figures and Captions

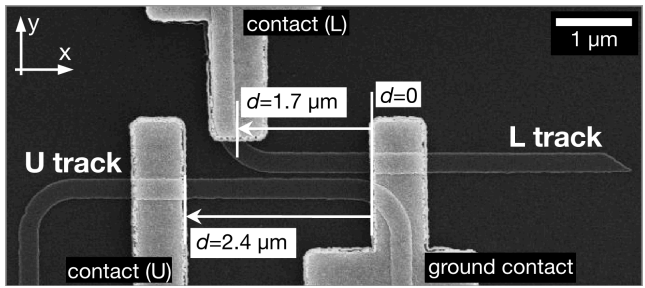

**Figure 1.**

SEM image of the two adjacent SV nanotracks (L and U tracks). The three electrical contacts (L, U, and ground) can also be seen. The coordinate $d$ is the distance along either track measured from the edge of the ground contact.

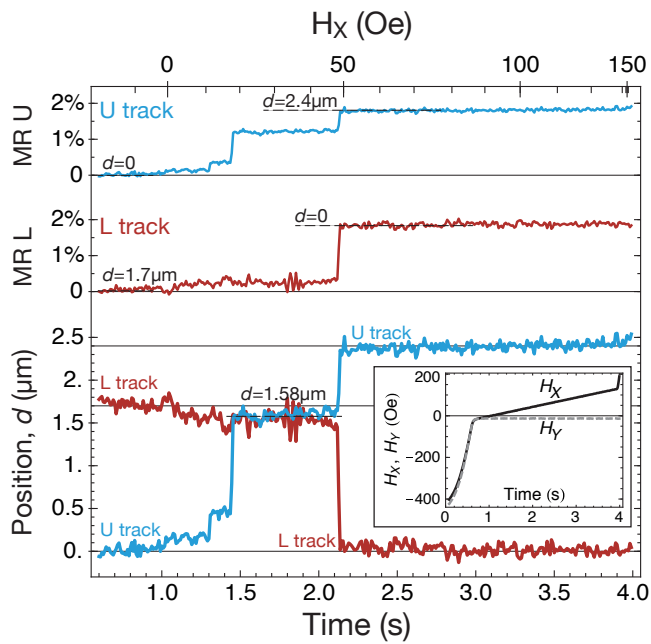

**Figure 2.**

(color online) Measurement of a DW induced pinning event. MR levels of both U (top curve) and L (middle curve) tracks, and calculated DW positions (bottom curve) versus time and $H_X$. Both DWs depin simultaneously from $d$=1.58μm. **inset.** External field components, $H_X$ and $H_Y$, versus time.

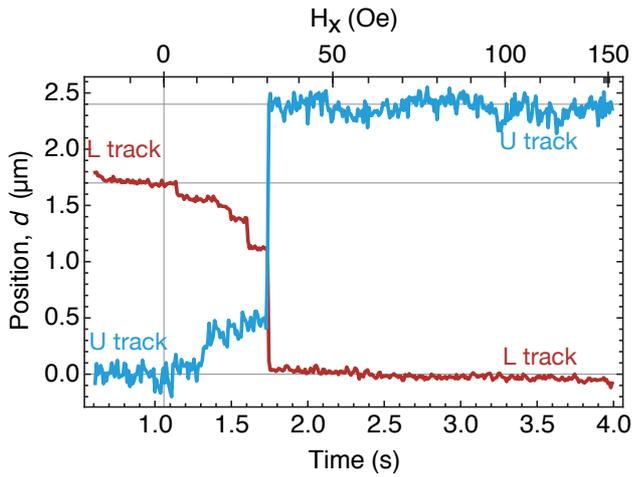

**Figure 3.**

(color online) Measurement of a DW induced depinning event. The positions of the DWs in both tracks are shown versus time and $H_X$, showing a simultaneous depinning of both DWs from distant pinning centers.

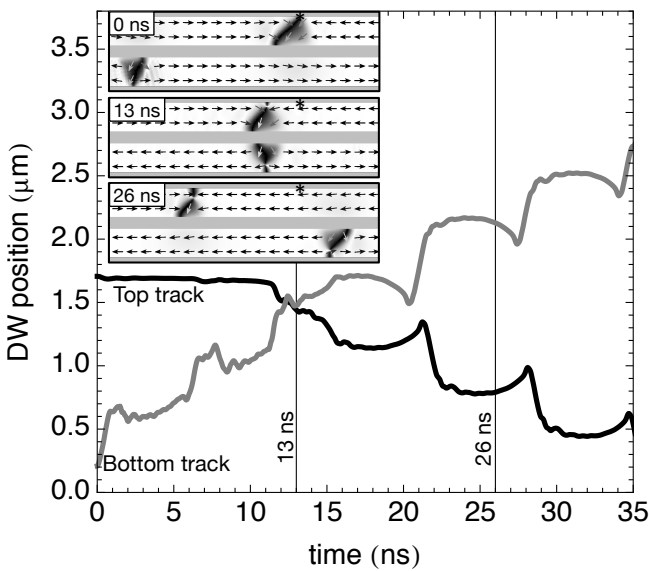

**Figure 4.**

Micromagnetic simulation of travelling DW induced depinning. Two magnetic tracks were simulated with a pinned DW in the top track and a travelling DW in the bottom, at constant field $H_0$=29 Oe (details in text). **Plot:** Position of both DWs versus time. **Inset:** Three normalized magnetization (*m*) snapshots (arrows follow *m*, the grayscale represents the vertical *m* component, void space in light gray), before (**t=0 ns**), during (**t=13 ns**), and after (**t=26 ns**) the passage of the travelling DW. The width of the tracks is 260 nm. The * marks the position of the pinning notch.